\documentstyle[11pt,newpasp,twoside,epsf]{article}
\def\edcomment#1{\iffalse\marginpar{\raggedright\sl#1\/}\else\relax\fi}
\reversemarginpar

\begin{document}
\title{Stellar Rotation from GAIA Spectra}
\author{Andreja Gomboc}
\affil{University of Ljubljana, Department of Physics, Jadranska 19, 1000 Ljubljana, Slovenia, 
andreja.gomboc@fmf.uni-lj.si}

\begin{abstract}
Stellar rotation influences our understanding of 
stellar structure and evolution, binary systems, clusters etc. and therefore
the benefits of a large and highly accurate database on stellar rotation, obtained by GAIA, will 
be manifold.

To study the prospects of GAIA measurement of projected rotational velocities $v_{rot} \sin i$,
we use synthetic stellar spectra to simulate the determination of $v_{rot} \sin i$ at different 
resolutions (R=5000 - 20000) and S/N (10-300).
Results on the accuracy of $v_{rot} \sin i$, presented here, show that
GAIA will be capable  
to measure also low rotational velocities ($\sim $ 10 km/s), provided that the resolution is higher than
10.000, or preferably even higher.
\end{abstract}

\section{Introduction}

At present there are approximately 20.000 stars with measured $v_{rot}\sin i$ (Glebocki, Gnacinski and
Stawikowski 2000). 
Sorting them by stellar type and rotational velocity (Munari 2002, Soderblom 2001) shows  that
early type stars (O, B, A, early F) have high rotational velocities (50-400 km/s) and late type stars 
(late F, G, K, M) are slow rotators (in majority with $v_{rot}\sin i<$50 km/s), 
which is mainly attributed to the fact that late type stars have convective 
envelopes, while early type stars do not. In view of measuring stellar rotational velocities with GAIA, it
is necessary to estimate the accuracy of measured $v_{rot}\sin i$, which helps us in answering the question, whether
slow rotators can be discerned from non-rotators with GAIA.

\section{Scientific questions related to stellar rotation}
Data obtained by GAIA with millions and millions of stellar spectra and measured rotational 
velocities will certainly crucially contribute to the current know\-ledge on stellar rotation and its influence 
on various aspects of stellar physics. To shortly name a few (Maeder and Meynet 2000, Soderblom 2001, Patten and Simon 1996):
\begin{itemize}
\item
The effect of stellar rotation on stellar structure and evolution: rotation of a star influences its structure, 
luminosity, position on the HR diagram, life time etc. 
Rotation can via rotationally induced mixing 
lead to He and N enrichment, chemically peculiar stars, 
it can cause turbulence, influence stellar winds and spots. 
\item
The impact of differential rotation on stellar structure, surface phenomeno\-logy, 
models for generating magnetic fields etc.
\item
Stellar rotation as the indicator of age (for low mass ZAMS stars) through the magnetic breaking and the indicator 
of photospheric activity.
\item
In binary systems: stellar rotation is important in understanding the number of questions, such as the question of 
synchronization between rotational and orbital periods and the effectivness of tidal energy dissipation.
\item
Rotational velocity as an indicator of angular momentum of solar type stars and how it is affected by 
the presence of massive distant planets.
\item
Stellar rotation as an indicator of age, total mass and binding energy in open clusters and as a clue to test various 
theories of fragmentation of the parent cloud.
\item
The orientation of rotational axis - are they really randomly oriented and therefore not connected with the 
rotation of the Galaxy, which is the question that only wide GAIA statistics can answer.
\end{itemize}

\section{Simulating GAIA accuracy of $v_{rot}\sin i$}

The important issue of GAIA contribution to stellar rotation physics is the accuracy of obtained  $v_{rot}\sin i$.
Here are presented results of simulating the determination of rotational velocities on synthetic stellar 
spectra. 
We used Kurucz synthetic stellar spectra database in the GAIA wavelength region.

To start the simulation we choose the spectrum with known T, Z, log g, v$_{rad}$and original $v_{rot}\sin i$,  
artificially add Poison distributed noise and afterwards fit it with noise-free spectra with various $v_{rot}\sin i$. 
As the best fitting spectrum we take the spectrum with 
minimum $\chi^2$ and repeat the test N times. The accuracy is determined as the standard 
deviation error between the original and recovered $v_{rot}\sin i$:
\begin{displaymath}
\sigma^2={{1\over N}\, \sum_{i=1}^N{\Bigl[v_i(orig)-v_i(rec)\Bigr]^2}}.
\end{displaymath} 

We used synthetic spectra of four star types: late G type giant (T=4750, 
Z=-0.5, log g=1.0), K dwarf  
(T=4750, Z=-0.5, log g=4.5), G type giant (T=5500, Z=-0.5, log g=2.0) 
and early F type star (T=7250, Z=0.0, log g=4.5), 
all with $v_{rot}\sin i$=10 km/s at different resolutions: R=5000, 8615, 17230 and 20000. 
Noise added corresponds to signal to noise ratio in the range of S/N=10-300 and the number of trials is  N=5000.
\begin{figure}
\plotfiddle{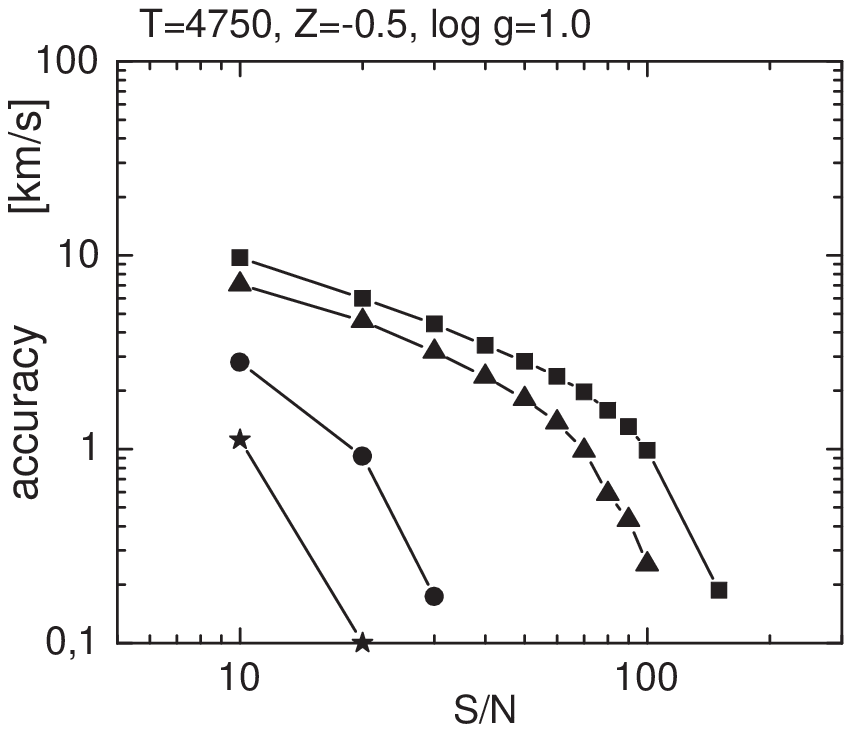}{1.7cm}{0}{65}{55}{-177}{-70}
\plotfiddle{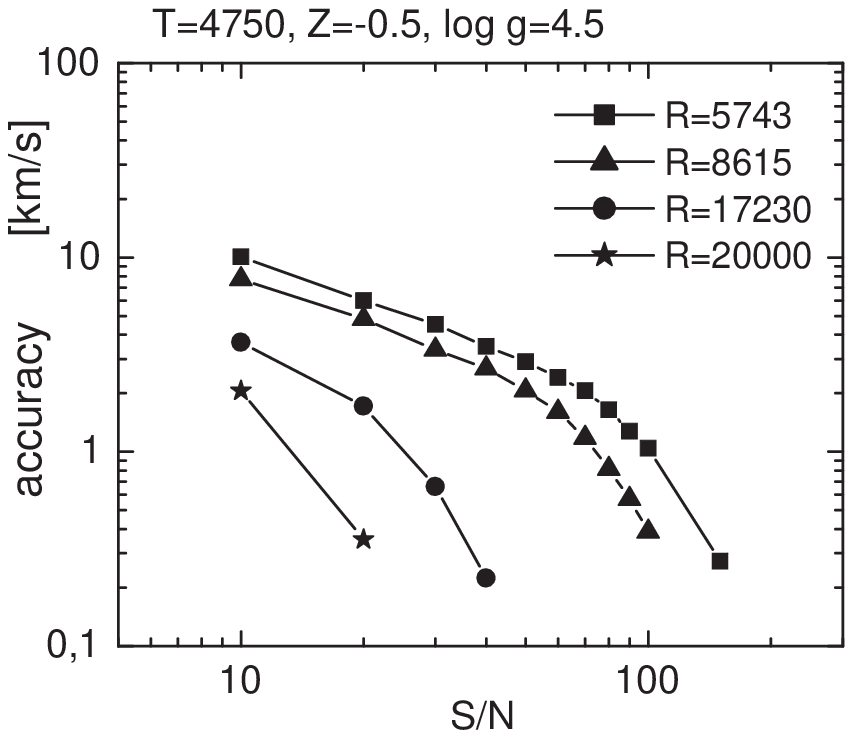}{1.7cm}{0}{65}{55}{-15}{-10}
\plotfiddle{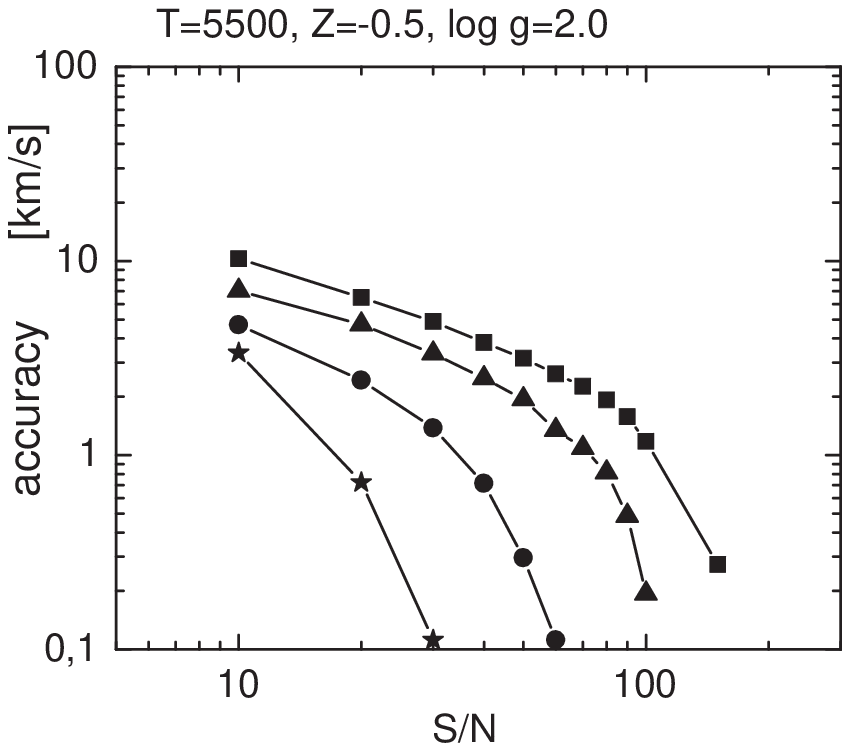}{1.7cm}{0}{65}{55}{-175}{-70}
\plotfiddle{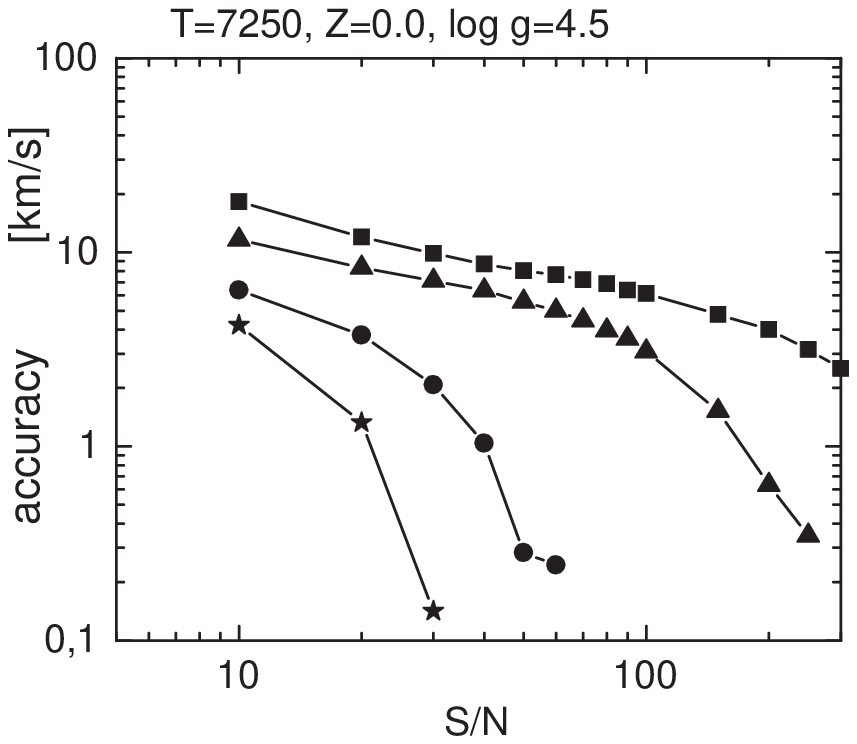}{1.7cm}{0}{65}{55}{-15}{-9}
\caption{The simulated accuracy of $v_{rot}\sin i$ for four types of stars at resolutions R=5743, 8615, 17230, 20000 
and as a function of S/N ratio. Other stellar parameters (T, Z, log g, v$_{rad}$) are presumed to be exactly known, i.e. 
the same in the original and fitting spectra.}
\end{figure}
\begin{figure}
\plotfiddle{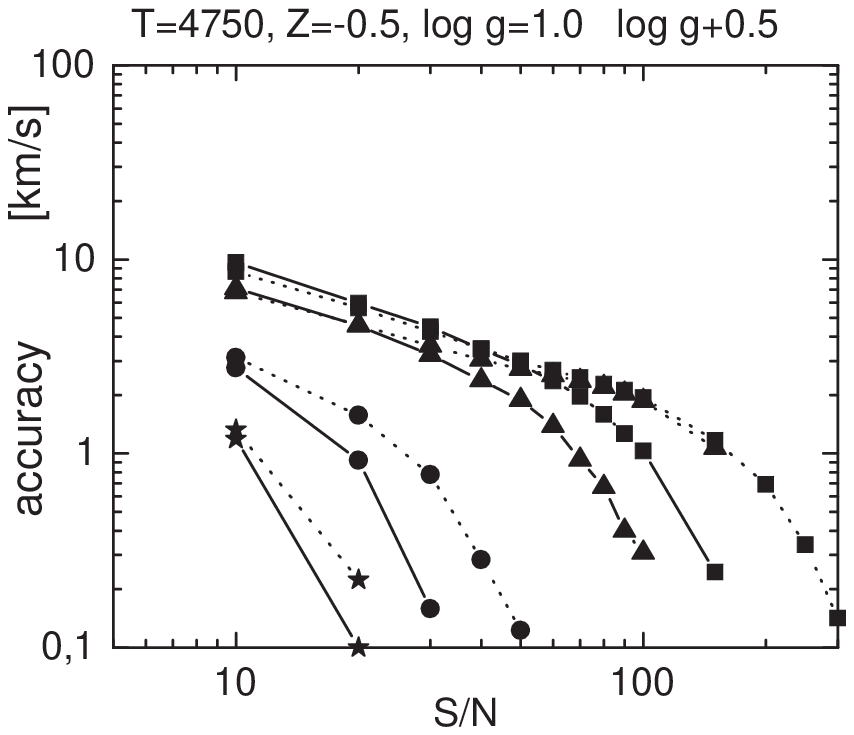}{1.7cm}{0}{65}{55}{-177}{-70}
\plotfiddle{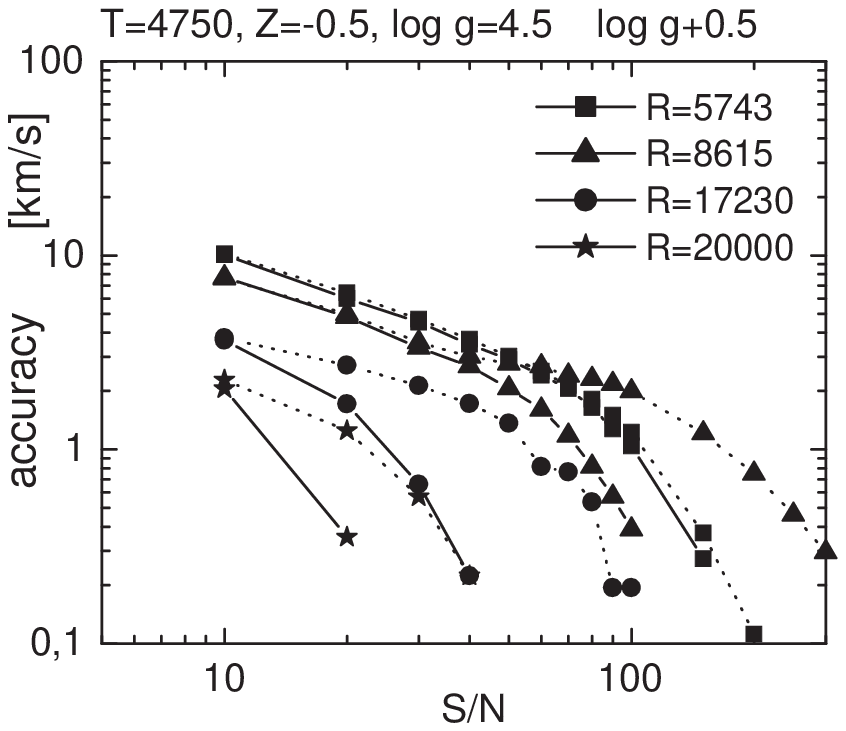}{1.7cm}{0}{65}{55}{-15}{-10}
\plotfiddle{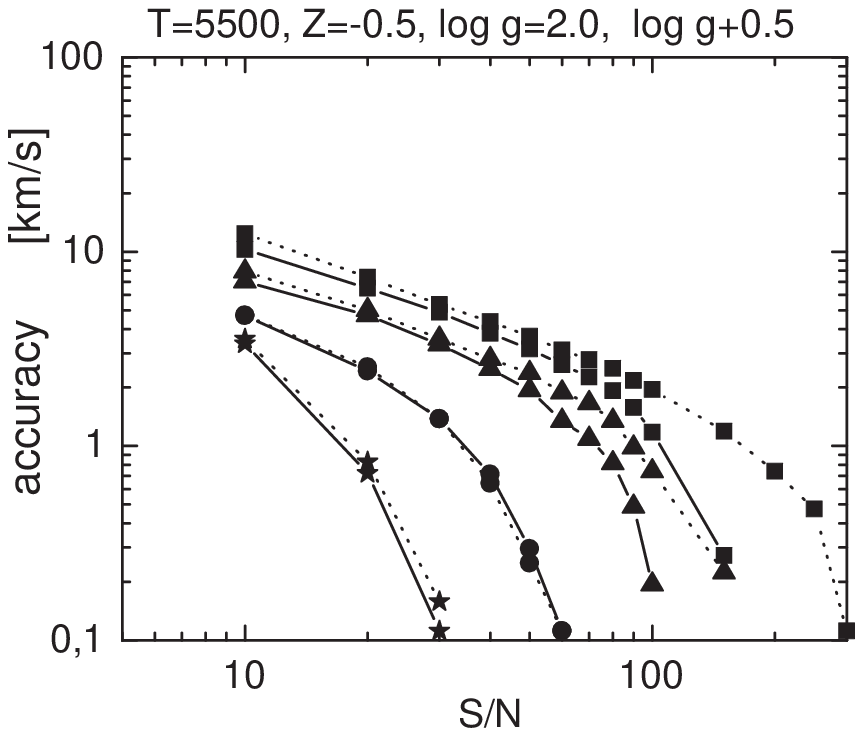}{1.7cm}{0}{65}{55}{-179}{-70}
\plotfiddle{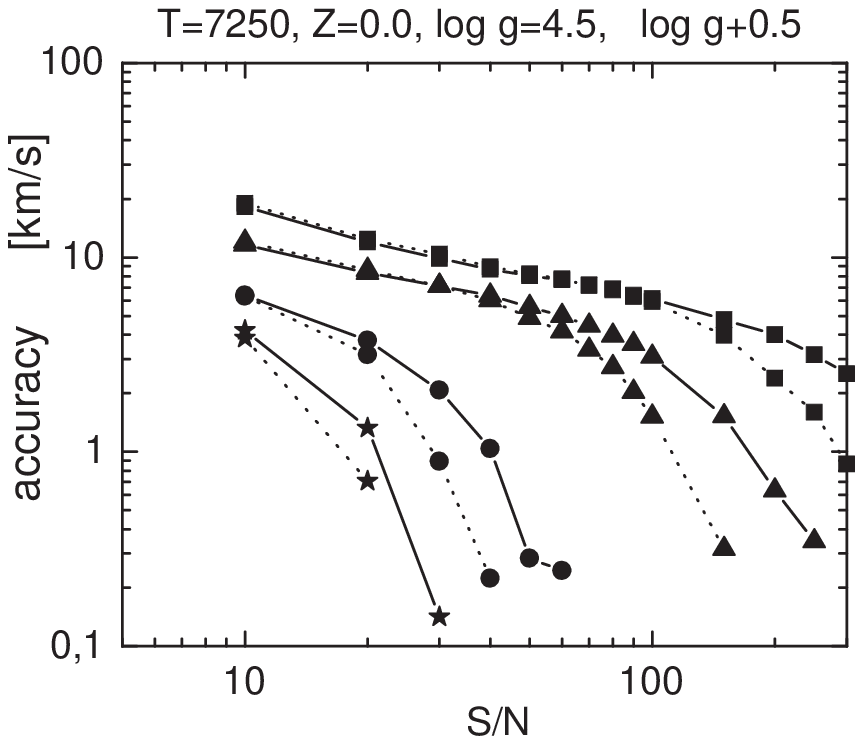}{1.7cm}{0}{65}{55}{-15}{-9}
\caption{The same as Fig. 1. Dotted lines show the accuracy obtained if log g of fitting spectra 
is offset from its true value by 0.5.}
\end{figure}

As the fitting spectra we first use the spectra with stellar parameters (T, Z, log g, v$_{rad}$) exactly the same as in the 
original spectrum, so that they differ only in $v_{rot}\sin i$. The obtained accuracy in such simulations is shown on 
Figure 1: as expected the accuracy is much 
better for high resolution and improves with better signal to noise. It should be stressed though, that this is 
the most ideal case, since other stellar parameters are precisely known. 

Due to possible uncertainties in these, the accuracy of rotational
velocity will generally become worse. To estimate the influence of those uncertainties on the accuracy of obtained 
rotational velocity, we performed the same simulations by fitting the original spectrum with 
spectral templates that had one of the parameters  (T, Z, log g or v$_{rad}$) offset from its true value.

The least crucial factor turns out to be the uncertainty in log g. The dotted curves in Figure 2 show the 
accuracy of rotational velocity if fitted by spectra having log~g offset for 0.5 from its original value.
\begin{figure}
\plotfiddle{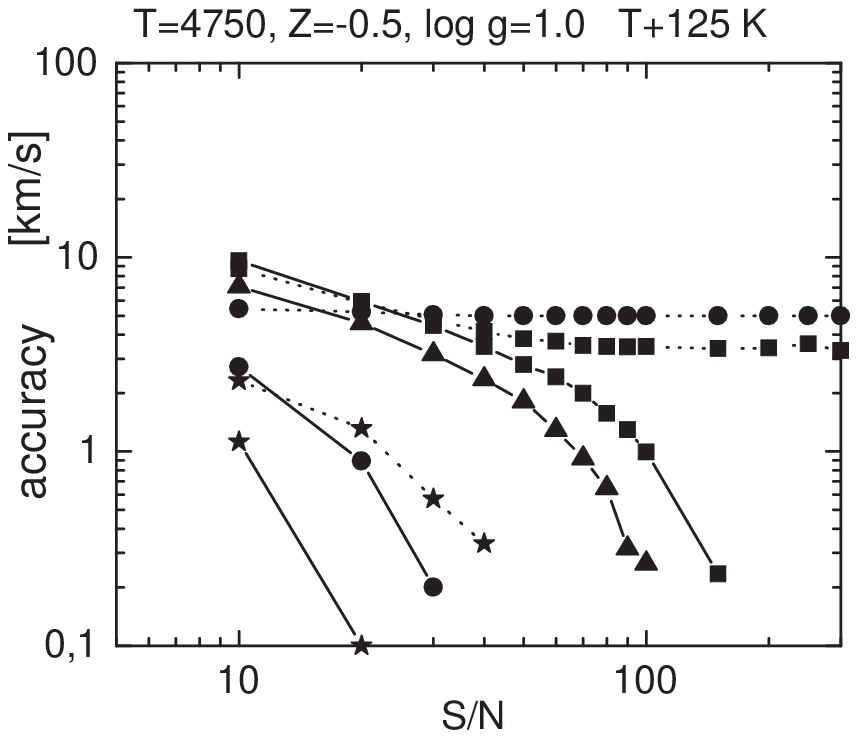}{1.7cm}{0}{65}{55}{-177}{-70}
\plotfiddle{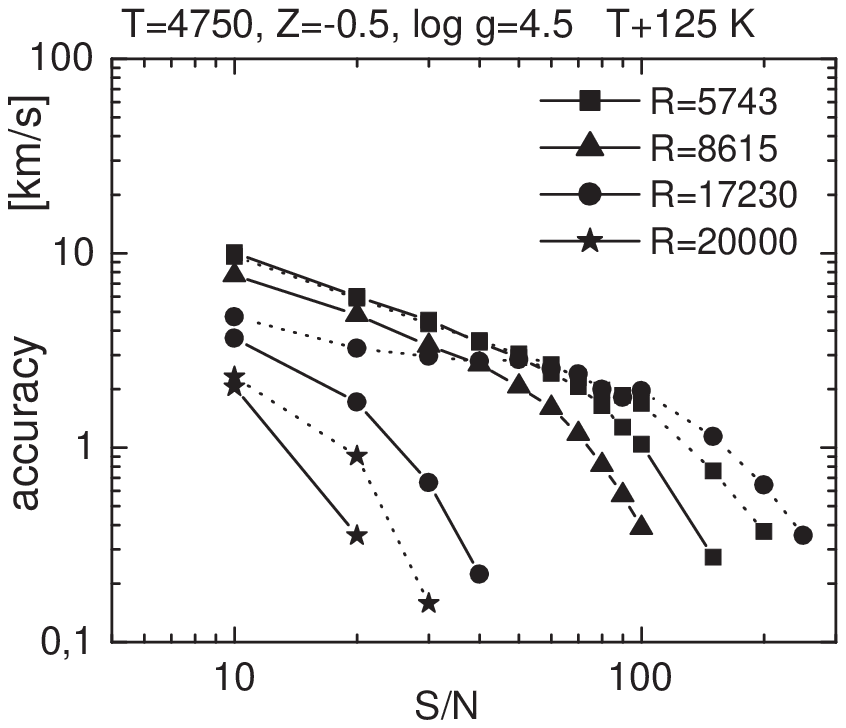}{1.7cm}{0}{65}{55}{-15}{-10}
\plotfiddle{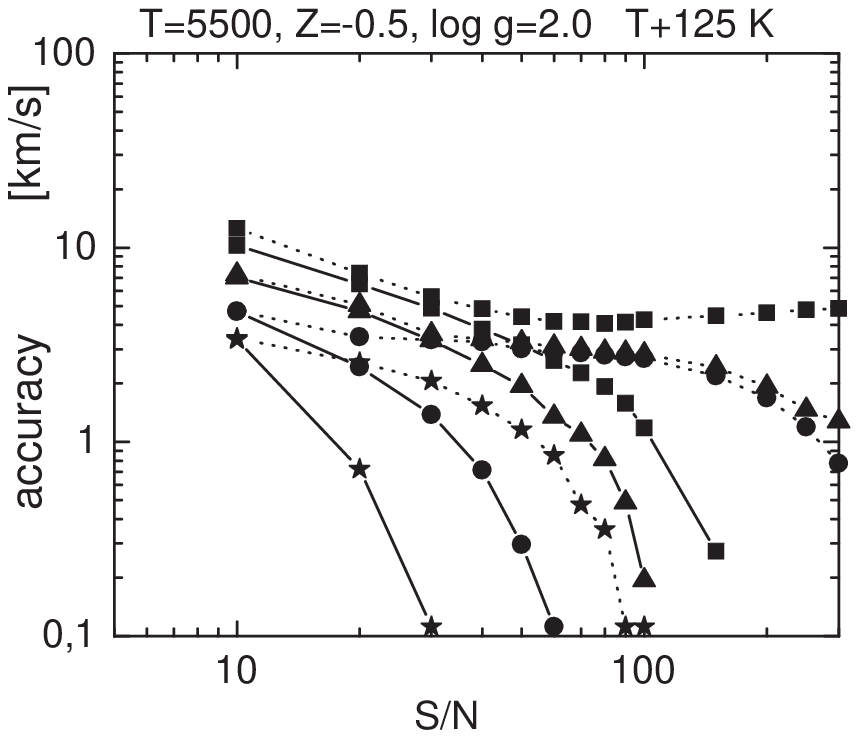}{1.7cm}{0}{65}{55}{-177}{-70}
\plotfiddle{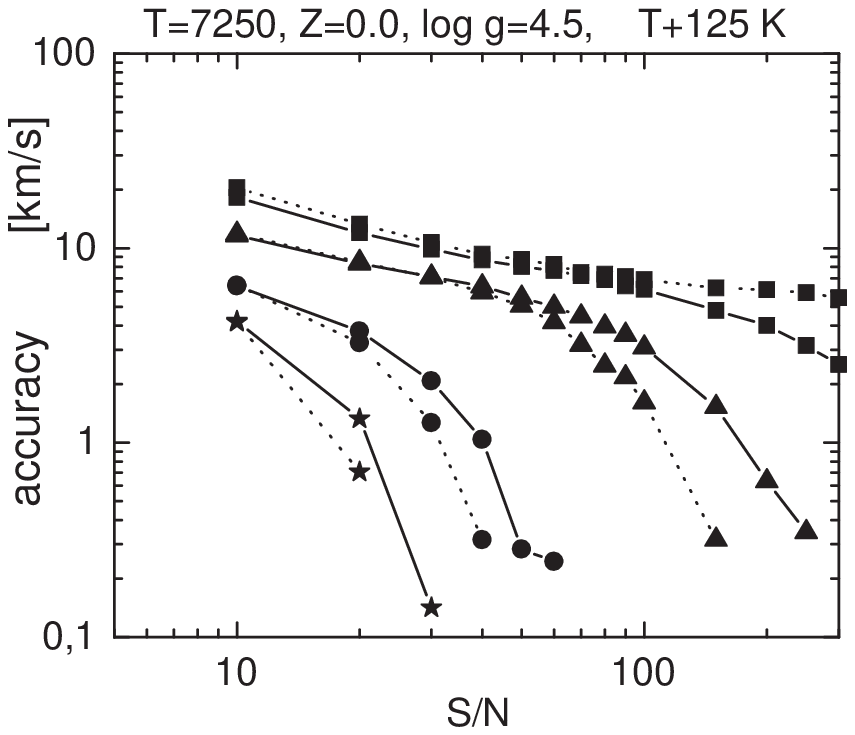}{1.7cm}{0}{65}{55}{-15}{-9}
\caption{The same as Fig.~1. Dotted lines show the accuracy obtained if T of the fitting spectra 
is offset from its true value by 125~K.}
\end{figure}

More crucial is an accurate temperature. The error of 250 K in some cases keeps the rotational velocity 
error at 5-10 km/s, not improving with S/N, while the error of 125~K (Figure 3) is small enough that at least at high 
resolutions the accuracy is better than 1~km/s at high S/N.
\begin{figure}
\plotfiddle{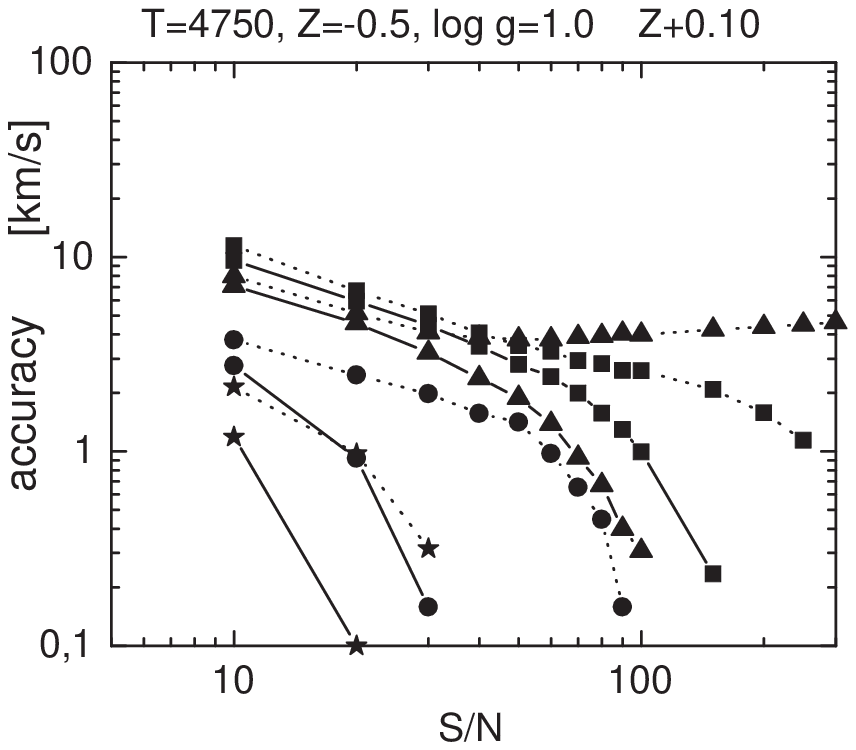}{1.7cm}{0}{65}{55}{-177}{-70}
\plotfiddle{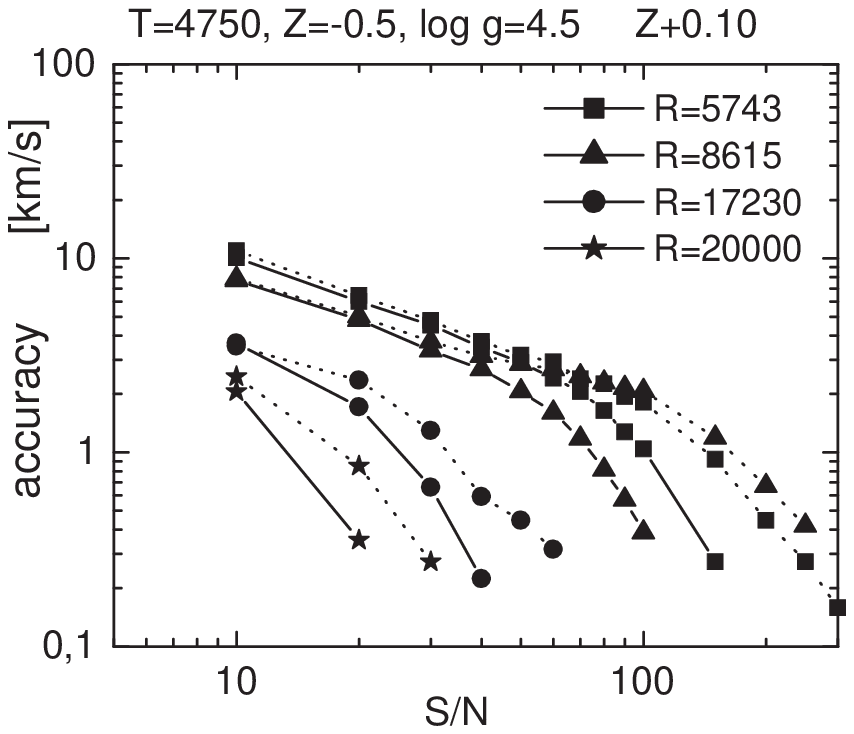}{1.7cm}{0}{65}{55}{-15}{-10}
\plotfiddle{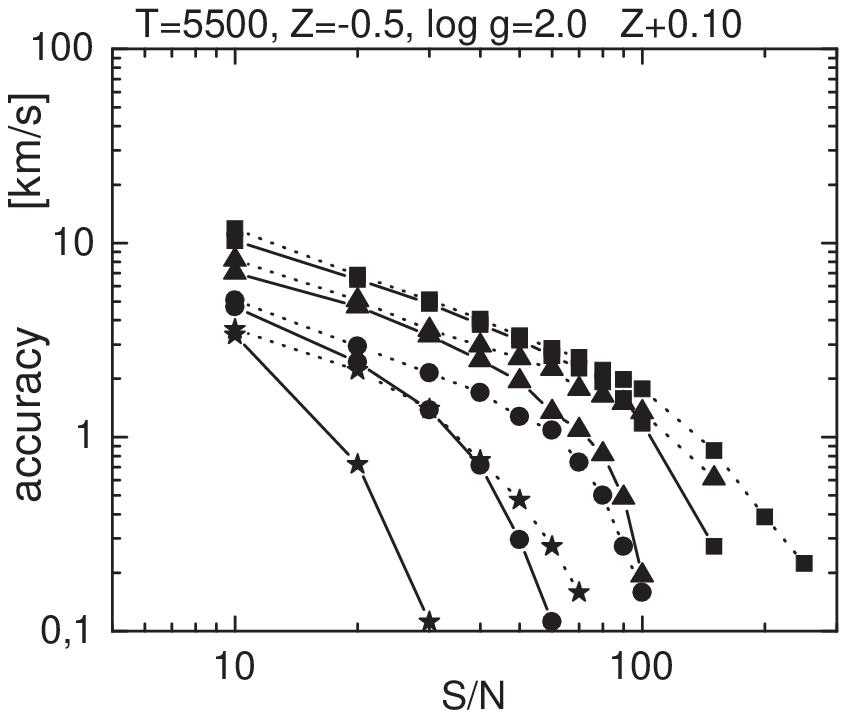}{1.7cm}{0}{65}{55}{-177}{-70}
\plotfiddle{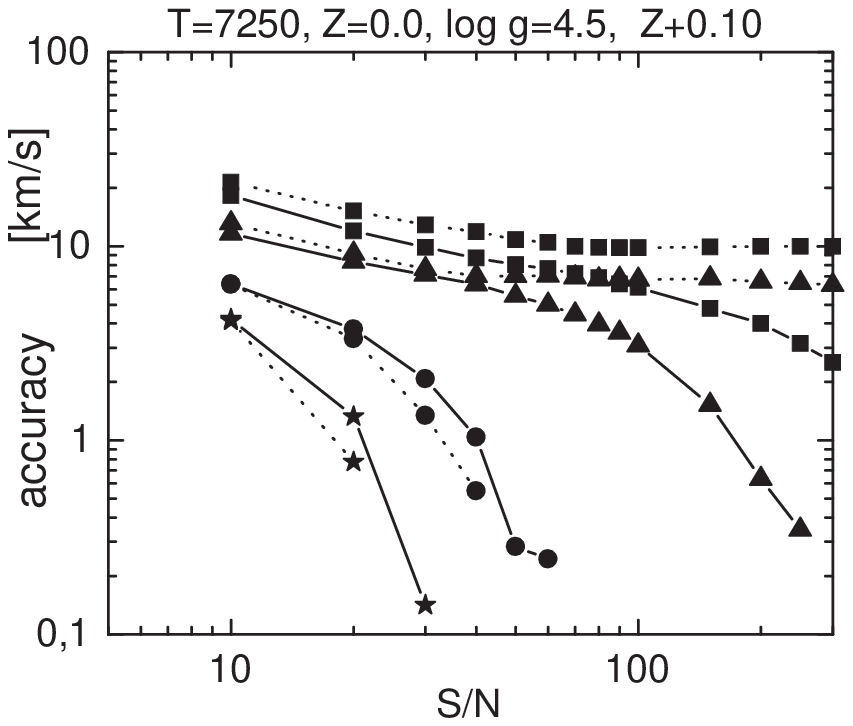}{1.7cm}{0}{65}{55}{-15}{-9}
\caption{The same as Fig. 1. Dotted lines show the accuracy obtained if Z of fitting spectra 
is offset from its true value by 0.1.}
\end{figure}

One of crucial factors is an accurate metallicity. The error of 0.5 in metallicity leads to inaccurate rotational velocities 
(error $>$10 km/s), the smaller error of 0.25 is about a factor of 2 better 
and 0.1 is quite good for high resolutions (Figure 4). Note that the accuracy of the 
rotational velocity does not seem to suffer so much if the metallicity is underestimated than if it is overestimated by 
the same amount.
\begin{figure}
\plotfiddle{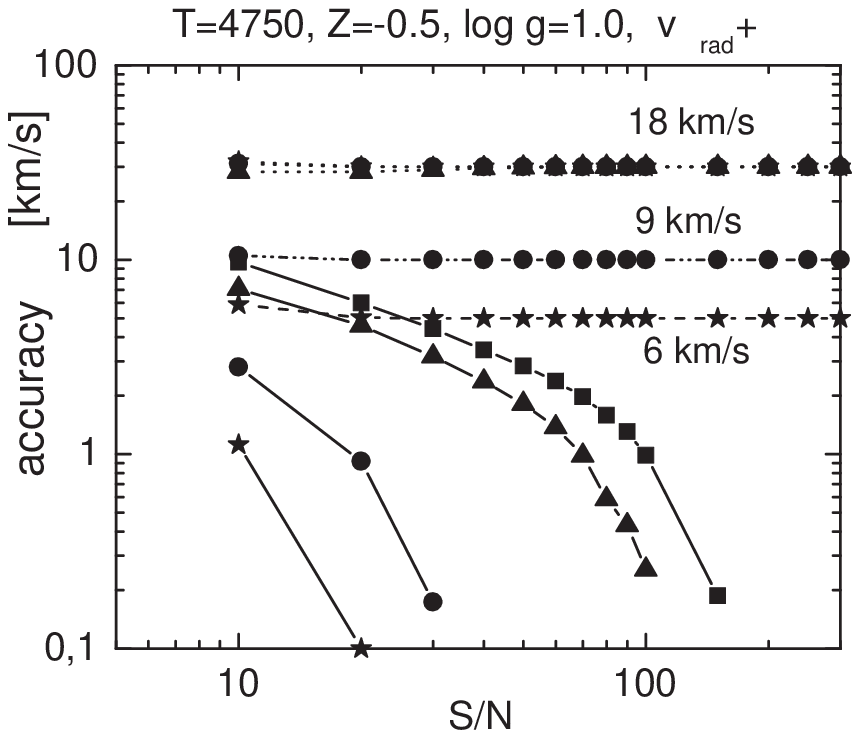}{1.7cm}{0}{65}{55}{-177}{-70}
\plotfiddle{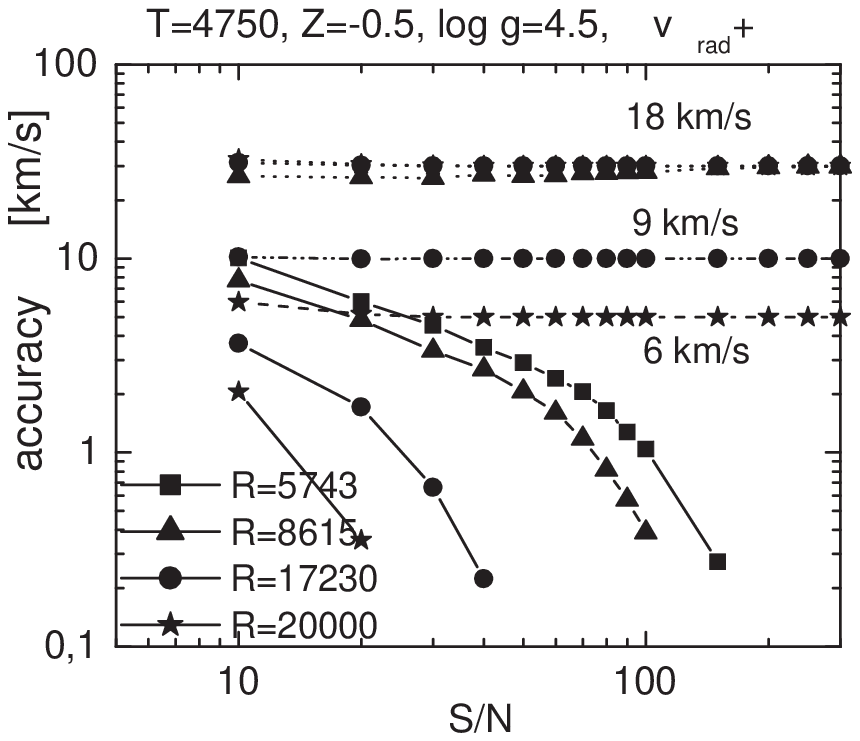}{1.7cm}{0}{65}{55}{-15}{-10}
\plotfiddle{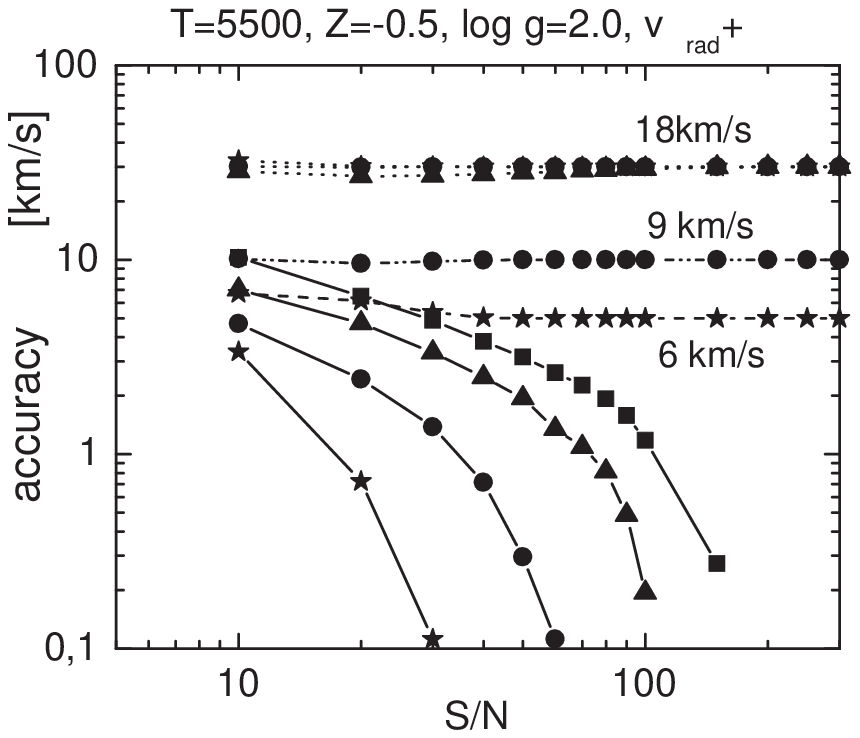}{1.7cm}{0}{65}{55}{-176}{-70}
\plotfiddle{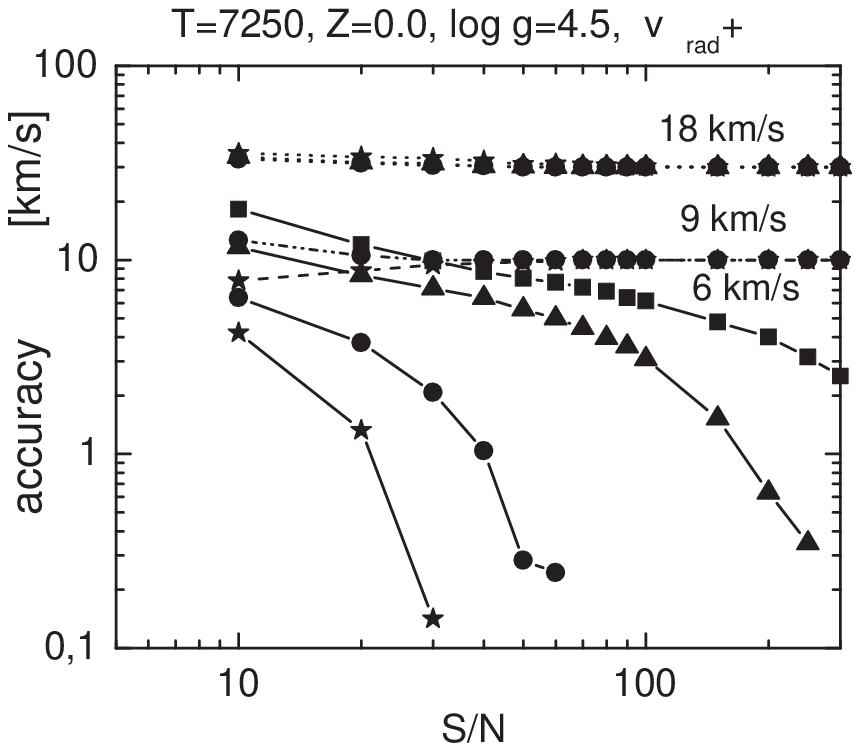}{1.7cm}{0}{65}{55}{-15}{-10}
\caption{The same as Fig. 1. Dashed and dotted lines show the accuracy obtained if v$_{rad}$ of fitting spectra 
is offset from its true value by (from top to bottom): 18, 9 and 6 km/s.}
\end{figure}

Crucial is also the accuracy of the obtained radial velocity. As shown in Figure 5, the error in radial 
velocity of 10 km/s leads to an error in rotational velocity of more than 10 km/s. If the radial velocity error
is 5~ km/s the error of rotation velocity is about 2 times smaller.

\section{Conclusion}
Table 1 gives the summary of stellar rotation accuracy obtained by simulations based on synthetic 
stellar spectra for different resolution and signal to noise ratio in the continuum. 
Columns 3-7 report the rotational velocity errors if the template used to recover the spectrum is ideal or 
mismatched for a given amount in temperature, metallicity, surface gravity or radial velocity.
\begin{table}
\caption{Estimated accuracy of $v_{rot}\sin i$ (in km/s):}
\begin{tabular}{crrcrrrr}
&&&&&&&\\
\tableline
&&&&&&&\\ [-5pt]
&\multicolumn{1}{c}{R}
&\multicolumn{1}{c}{S/N}
&\multicolumn{1}{c}{ideal}
&\multicolumn{1}{c}{$\Delta$log g=0.5}
&\multicolumn{1}{c}{$\Delta$T=125 K}  
&\multicolumn{1}{c}{$\Delta$Z=0.1}  
&\multicolumn{1}{c}{$\Delta$v$_{rad}$=5 km/s}  
\\ [4pt]
\tableline
&&&&&&&\\ [-5pt]
 & 5000 & 10 & 10-20      &  10-20 & 10-20 & 10-20 &  \\
 &  & 100 & 2-7&     2-8      &  2-10 & 2-15 &  \\
 & 10000 & 10 & 7-12     & 7-12   & 7-12 & 8-13 &   \\
 &  & 100 & 0.2-3    & 1-3  & 2-5 & 2-8 &   \\
 & 20000 &  10 & 1-4 & 1-4 &  2.5-4  & 1-4 &  5-10 \\ 
 &  & 100 & $<$0.1      &  $<$0.1  & $<$0.1  & $<$0.1  & 5-10 \\ [4pt]
\tableline
\end{tabular}
\end{table}

The results we obtained can be used only as a rough indicator of GAIA capabilities. We used synthetic stellar 
spectra without allowing for modeling uncertainties, chemical composition peculiarities etc. The results of 
simulations show that effects of combined errors in two of the template parameters (e.g. T and Z, or Z and log g)
on the accuracy of rotational velocity are not always easily predictable. 
One should also be aware that the crowding of stellar spectra in GAIA focal plane may smear out some line profile
details, which may be crucial in determining the rotational broadening of spectral lines.
Nevertheless, we believe that measuring even small rotational velocities with GAIA is not out of reach, provided that 
the resolution is better than 10000, or preferably better than 15000.

\end{document}